
\input phyzzx
\FRONTPAGE
\line{\hfill}
\vskip.3cm
\centerline{{\bf TOWARDS A NONSINGULAR UNIVERSE}\footnote{a}{Work
supported in part by the U.S. Department of Energy under grant DE-
AC02-03130, Task A.}}
\vskip.7cm
\centerline{ROBERT H. BRANDENBERGER}
\centerline{{\it Department of Physics}}
\centerline{{\it Brown University, Providence, RI 02912, USA}}
\vskip1.5cm
\centerline{\bf INTRODUCTION}
\par
A unified theory of all forces should be nonsingular.  In such unified
theories, Einstein's theory of general relativity will be a very low
curvature effective theory.  At larger curvatures, new terms will
become important.  The question then arises as to whether it is
possible to construct an effective action for gravity which leads to a
nonsingular theory.

In this work$^{1)}$ we construct an effective action for gravity in
which all homogeneous and isotropic solutions are nonsingular.  In
particular, there is neither a big bang nor a big crunch.  The theory
is a higher derivative modification of Einstein's theory constructed
in analogy to how the action for point particle motion in special
relativity is obtained from Newtonian mechanics.  Preliminary
work$^{2)}$ indicates that our construction provides a theory in which
anisotropies decrease at large curvatures, and in which the inside of
a black hole is singularity free.

As is well known, the Penrose-Hawking singularity theorems$^{3)}$
prove that space-time manifolds in general relativity are geodesically
incomplete.  The key assumptions which are used in the proofs of these
theorems is that the action for gravity is the unmodified Einstein
action, and that matter satisfies the energy dominance condition
$\varepsilon > 0$ and $\varepsilon + 3 p \ge 0$, where $\varepsilon$ and $p$
are energy density and pressure respectively.

In general, the singularity theorems do not provide any information on
the nature of the singularity.  However, in well known examples such
as a collapsing homogeneous and isotropic Universe or the
Schwarzschild black hole, at the singularity at least one of the
curvature invariants blows up.

Solutions to the ``singularity problem" of general relativity such as
quantum gravity or string theory all lead to low energy effective
actions which include higher derivative terms.  As a guide towards
constructing an acceptable unified theory we will ask which class of
higher derivative effective actions for space-time gives a nonsingular
Universe.

Our goal is to construct a theory in which all curvature invariants
are bounded and which in addition is geodesically complete.  This
formidable problem can be reduced substantially by invoking the
``Limiting Curvature Hypothesis" $^{4)}$, according to which one
\item{i)} finds a theory in which a small number of specially chosen
invariants are explicitly bounded, and
\item{ii)} when these invariants approach their limiting values, a
definite nonsingular solution (namely de Sitter) is taken on.

\noindent
As a consequence of the limiting curvature hypothesis, automatically
all invariants are bounded, and space-time is geodesically complete in
its asymptotic regions.

The Limiting Curvature Hypothesis has interesting consequences for
Friedman models and for spherically symmetric space-times$^{5)}$.  A
collapsing Universe will not end up in a big crunch, but will approach
a contracting de Sitter Universe $(k = 0)$ or a de Sitter bounce $(k
= 1)$ followed by re-expansion (see Fig. 1).  For a spherically
symmetric vacuum solution, there will be no singularity inside the
Schwarzschild horizon; instead, a de Sitter Universe will be reached
(see Fig. 1).
\endpage
\line{\hfill}
\vskip11.0cm
\noindent{\bf Figure 1.}
Penrose diagrams for a collapsing Universe (left)
and for a black hole (right) in Einstein's theory (top) and after
implementing the Limiting Curvature Hypothesis (bottom).  Wavy lines
denote a singularity, the symbols $C$, $DS$ and $E$ stand for
collapsing phase, de Sitter period, and expanding phase respectively,
and $H$ denotes the Schwarzschild horizon.
\vskip0.4cm
\centerline{\bf CONSTRUCTION}

We shall construct a nonsingular Universe by considering a special
class of higher derivative gravity models$^{1)}$.  Higher derivative
gravity theories arise as effective actions at high curvatures in many
circumstances, {\it e.g.}, in quantum gravity, when considering quantum
fields in an expanding Universe, or in the low energy limit of string
theory.  It is hence very reasonable to consider such theories.

We construct our special class of models in analogy with how the
action for point particle motion in special relativity can be obtained
from Newtonian mechanics$^{6)}$.  Starting point is the Newtonian
action for a point particle with world line $\undertext{x}(t)$ and mass
$m$
$$
S = m \int dt \, {1\over 2} \dot x^2 \, . \eqno\eq
$$
In this theory, there is no bound on the velocity $\dot x$.  We
construct a new theory by adding a Lagrange multiplier $\varphi$ which
couples to $\dot x^2$, the scalar quantity which is to be limited, and
give $\varphi$ a potential $V(\varphi)$.  The new action is
$$
S = m \int dt \, \left[ {1\over 2} \dot x^2 + \varphi \dot x^2 - V
(\varphi) \right] \, . \eqno\eq
$$
Provided that $V (\varphi) \sim \varphi$ for $|\varphi | \rightarrow
\infty$, the constraint equation ({\it i.e.}, the variational equation
with respect to $\varphi$) ensures that $\dot x$ is bounded.  In order
to obtain the correct Newtonian limit for small $\dot x$ ({\it i.e.},
small $\varphi$), $V(\varphi)$ must be proportional to $\varphi^2$ for
$| \varphi | \rightarrow 0$.  Up to factors of 2, the simplest
potential which satisfies the above asymptotic conditions is
$$
V (\varphi) = {2 \varphi^2 \over{1 + 2 \varphi}} \, . \eqno\eq
$$
Eliminating the Lagrange multiplier using the constraint equation and
substituting the result into the action, yields the point particle
action
$$
S = m \int \, dt \sqrt{1 - \dot x^2} \, \eqno\eq
$$
in special relativity.

The first step of implementing the Limiting Curvature Hypothesis
follows the above procedure.  As ``old theory" we take Einstein's
action, and from it we construct a new theory in which the Ricci
scalar $R$ is limited by introducing a Lagrange multiplier $\varphi_1$
with potential $V_1 (\varphi_1)$ which couples to $R$:
$$
S = \int d^4 x \sqrt{-g} \, [ R + \varphi_1 R + V_1 (\varphi_1) ]
\eqno\eq
$$
where $V_1 (\varphi_1)$ must satisfy the same asymptotic conditions as
the multiplier $\varphi$ in our special relativity example.

As a second step in implementing the Limiting Curvature Hypothesis, we
must ensure that as $R$ tends to its limiting values, space-time
becomes de Sitter.  It is possible to accomplish this by once again
using a Lagrange multiplier construction.  We find an invariant $I_2
(R^{\alpha \beta \gamma \delta})$ with the properties that $I_2 \ge 0$
and $I_2 = 0$ if and only if space-time is de Sitter.  We couple a new
Lagrange multiplier $\varphi_2$ to $I_2$ and choose its potential $V_2
(\varphi_2)$ such that as $| \varphi_2 | \rightarrow \infty \>\, V_2
(\varphi_2) \rightarrow {\rm const.}$  Hence, at large $|\varphi_2|$,
$I_2$ is driven to zero and de Sitter space results.

Thus, the new action is
$$
S = \int d^4 x \, \sqrt{-g} \left[ R + \varphi_1 R - \left( \varphi_2
+ {6\over{\sqrt{12}}} \varphi_1 \right) \, I_2^{1/2} + V_1 (\varphi_1)
+ V_2 (\varphi_2) \right] \eqno\eq
$$
where the coupling of $\varphi_1$ to $I_2^{1/2}$ is introduced to simplify
the final equations.  For a homogeneous and isotropic metric, the
invariant $I_2$ can be chosen to be
$$
I_2 = 4 R_{\mu \nu} R^{\mu\nu} - R^2 \eqno\eq
$$
which is positive for general metrics and vanishes (among homogeneous
and isotropic metrics) only for de Sitter spaces.

A simple realization of our scenario uses the potentials
$$
\eqalign{V_1 (\varphi_1) & = 12 \, H_0^2 \, {\varphi_1^2\over{1 +
\varphi_1}} \, \left( 1 - {\ln (1 + \varphi_1)\over{1 + \varphi_1}}
\right) \cr
V_2 (\varphi_2) & = - \sqrt{12} \, H_0^2  \, {\varphi_2^2\over{1 +
\varphi_2^2}} \, . } \eqno\eq
$$
Applied to a spatially flat collapsing Friedmann-Robertson-Walker
metric with scale factor $a(t)$ and Hubble expansion rate $H (t) =
\dot a (t)/a (t)$, the following variational equations result
$$
\eqalign{ & H^2 = {1\over 12} \, V_1^\prime \cr
& \dot H = - {1\over{\sqrt{12}}} \, V^\prime_2 \cr
& 3 (1 - 2 \varphi_1) H^2 + {1\over 2} \, (V_1 + V_2) =
{6\over{\sqrt{12}}} \,  H \,  (\dot \varphi_2 + 3 H \varphi_2) \, . } \eqno\eq
$$

{}From the construction it is already obvious that no singularities
arise for trajectories once they reach the asymptotic region $|
\varphi_2 | \rightarrow \infty$.  Equations (9) must be investigated
explicitly to show that no singularities arise for finite values of
$\varphi_2$, and that trajectories are either periodic about
$(\varphi_1 \, , \varphi_2) = ( 0, 0)$, or else asymptotically
approach $| \varphi_2 | \rightarrow \infty$.

It has been verified by analytic phase diagram analysis and by
numerically solving (9) that indeed no singular points arise in the
entire $( \varphi_1 \, , \varphi_2)$ plane and that all nonperiodic
solutions are asymptotically de Sitter (for more details and figures
the reader is referred to Refs. 1 \& 7).

It is interesting to include matter in our consideration.  Thus, we
consider an action
$$
S_{\rm full} = S + S_m \eqno\eq
$$
where $S$ is the gravitational action (6) and $S_m$ is the matter
action for an ideal gas with pressure $p = 0$ (dust) or $p = {1\over
3} \rho$ (radiation).  In either case, it can be shown$^{7)}$ that the
asymptotic solutions $| \varphi_2 | \rightarrow \infty$ are unchanged.
In particular, the system remains singularity free.

This result implies that in our theory, gravity is asymptotically free
in the sense that the coupling between matter and gravity becomes
negligible at high curvature.

We can also study the variational equations which follow from (6) for
a spatially closed homogeneous and isotropic Universe.  The only
change compared to the spatially flat case is that the de Sitter
contraction is replaced by a de Sitter bounce.
\vskip0.4cm

\centerline{\bf SPECULATIONS}

If we apply our model (6) to an expanding Universe, we conclude that
it has emerged from an initial de Sitter phase, {\it i.e.}, from an
inflationary period.  This is not too surprising since many higher
derivative gravity theories yield inflation$^{8)}$.

However, what is new in our model is that gravity is asymptotically
free in the inflationary phase.  This might have important
consequences for the scalar metric perturbations produced during
inflation$^{9)}$.  Their amplitude might be suppressed compared to the
usual inflationary models.  There is the hope of having inflation
without the fine tuning problem of density perturbations.

For black holes we conjecture that as the black hole mass shrinks to a
mass $M_{\rm crit}$ when the Weyl curvature at the horizon reaches its
limiting value, Hawking radiation shuts off.  There are no loss of
unitarity problems associated with the final stages of black hole
evaporation.  Finite mass black hole remnants remain, and there is no
global charge violation by black holes.

In order for this last conjecture to be reasonable, we must first show
that in our theory black hole interiors are nonsingular.
\vskip0.4cm
\endpage
\centerline{\bf EXTENSIONS}

Consider the Schwarzschild metric
$$
ds^2 = \left( 1 - {2GM\over r} \right) dt^2 - \left( 1 - {2GM\over r}
\right)^{-1} dr^2 - \, (dy^2 + dz^2) \eqno\eq
$$
where $y$ and $z$ are local Cartesian coordinates on $S^2$.  Inside
the horizon, the metric can be approximated by
$$
\eqalign{ds^2 & \simeq {r\over{2GM}} dr^2 - {2 GM\over r} \, dt^2 \, -
 r^2 \, (dy^2 + dz^2) \cr
& = a^2 (\tau) \left[ d \tau^2 - e^{-2 \beta (\tau)} dx^2 - e^{\beta
(\tau)} \, (dy^2 + dz^2) \right] \, , } \eqno\eq
$$
where $\tau$ is a function of $r$ such that $a (\tau) = {1\over
\tau}$ and $r \rightarrow 0$ implies $\tau \rightarrow \infty$.  The
variable $x$ is a function of $t$.  The above analysis demonstrates
that the black hole interior is equivalent to a collapsing Kasner
Universe with two scale factors identified.

Hence, if we manage to show that the anisotropy decreases in a
collapsing Kasner Universe, we have also shown that the black hole
interior will approach de Sitter space, verifying the conjecture of
Fig. 1.

Obviously, the invariant $I_2$ of (7) is insufficient for anisotropic
models and for the black hole metric: $I_2$ vanishes identically for
the Schwarzschild metric.  However, there is a simple extension:
$$
I_2 = 4 R_{\mu \nu} R^{\mu\nu} - R^2 + C^2 \eqno\eq
$$
where $C^2 = C_{\alpha \beta \gamma \delta} C^{\alpha \beta \gamma
\delta}$, $C_{\alpha \beta \gamma \delta}$ being the Weyl tensor.  For
spherical symmetry $C^2 \geq 0$.  Hence, forcing $I_2 = 0$ implies
both $C^2 = 0$ (vanishing anisotropy) and $4R_{\mu\nu} R^{\mu\nu} -
R^2 = 0$ which means that the resulting isotropic space-time is de
Sitter$^{2)}$.

To simplify the algebra we can consider a simpler action$^{10,2)}$:
$$
S_A = \int d^4 x \, \sqrt{-g} \, (R + \varphi C^2 + V (\varphi) )
\eqno\eq
$$
with $V(\varphi) = 1 - \varphi^{-1/2}$ for $| \varphi | \rightarrow
\infty$.  It is not hard to obtain the variational equations for this
problem and to verify that for asymptotic solutions $| \varphi |
\rightarrow \infty$ the anisotropy decreases and de Sitter space is
approached.

Since these results should carry over to our system of (6) and (13),
we conclude that in our model, all homogeneous solutions will approach
de Sitter space, and that the black hole interior will be nonsingular.
\vskip0.4cm

\centerline{\bf CONCLUSIONS AND ACKNOWLEDGEMENTS}

We have constructed an effective action for gravity, based on a higher
derivative modification of Einstein's theory of general relativity, in
which all homogeneous solutions are nonsingular.  They are either
periodic about Minkowski space-time, or else they asymptotically
approach de Sitter space.  We have given some preliminary evidence
that in our model also singularities inside of black holes can be
avoided.

The results and ideas presented in this lecture are based on key ideas
by and joint work with V. Mukhanov.  I am grateful to him and to my
collaborators M. Mohazzab, A. Sornborger and M. Trodden.
\vskip0.4cm

\REF\one{V. Mukhanov and R. Brandenberger, {\it Phys. Rev. Lett.} {\bf
68}, 1969 (1992).}
\REF\two{R. Brandenberger, M. Mohazzab, V. Mukhanov, A. Sornberger and
M. Trodden, in preparation (1993).}
\REF\three{R. Penrose, {\it Phys. Rev. Lett.} {\bf 14}, 57 (1965);
\nextline
S. Hawking, {\it Proc. R. Soc. London} {\bf A300}, 182 (1967).}
\REF\four{M. Markov, {\it Pis'ma Zh. Eksp. Theor. Fiz.} {\bf 36}, 214
(1982); {\bf 46}, 342 (1987); \nextline
V. Ginsburg, V. Mukhanov and V. Frolov, {\it Zh. Eksp. Teor. Fiz.}
{\bf 94}, 1 (1988).}
\REF\five{V. Frolov, M. Markov and V. Mukhanov, {\it Phys. Lett.} {\bf
B216}, 272 (1989); \nextline
{\it Phys. Rev.} {\bf D41}, 383 (1990).}
\REF\six{B. Altshuler, {\it Class. Quant. Grav.} {\bf 7}, 189 (1990).}
\REF\seven{R. Brandenberger, V. Mukhanov and S. Sornborger, ``A
Cosmological Theory Without Singularities" Brown preprint BROWN-HET-891
(1993).}
\REF\eight{A. Starobinskii, {\it Phys. Lett.} {\bf B91}, 99 (1980).}
\REF\nine{V. Mukhanov, H. Feldman and R. Brandenberger, {\it Phys.
Rep.} {\bf 215}, 203 (1992).}
\REF\ten{B. Altshuler, in Proc. of the 1st Int. A.D. Sakharov
Conference, Moscow, May 1991, to be published in the proceedings
(Nova Science Publ., Inc., New York, 1992)}
\refout
\end